\documentclass{article} 
\usepackage{spconf,amsmath,graphicx}

\usepackage{enumitem}
\setlist{nosep, leftmargin=14pt}
\usepackage{amsfonts}
\usepackage{mwe} % to get dummy images
\usepackage{color}
\usepackage{multirow}
\usepackage{caption}
\usepackage{booktabs}
\usepackage{epsfig}
\usepackage{longtable}
\usepackage{rotating}
\usepackage{gensymb}
\usepackage{lscape}
\usepackage[colorlinks=true,
            linkcolor=red,
            citecolor=blue]{hyperref}
\graphicspath{{fig/}}
\usepackage{caption}
\usepackage{array, floatrow, tabularx, makecell, booktabs}%
\usepackage{amsmath}
\usepackage[utf8]{inputenc}
\begin{titlepage}
\title{Micro-CT Synthesis and Inner Ear Super Resolution via Generative Adversarial Networks and Bayesian Inference}

\name{ \normalsize Hongwei Li$^{*1, 2}$, Rameshwara Prasad$^{*2}$, Anjany Sekuboyina$^{2}$, Chen Niu$^{3}$, Siwei Bai$^{4, 5}$ 
{Werner Hemmer}$^{4, 5}$, and {Bjoern Menze}$^{1, 2}$} 
\address{ \normalsize 1. Department of Quantitative Biomedicine, University of Zurich \\ 
\normalsize 2. Department of Computer Science, Technical University of Munich \\ 
\normalsize 3. Department of Medical Imaging, First Affiliated Hospital of Xi’an Jiaotong University, China \\ 
\normalsize 4. Department of Electrical and Computer Engineering, Technical University of Munich \\ 
\normalsize 5. School of Bioengineering, Technical University of Munich\\
\normalsize Email: \{hongwei.li, bjoern.menze\}@uzh.ch}
\end{titlepage}
\begin{document}
\maketitle
\begin{abstract}
Existing medical image super-resolution methods rely on pairs of low- and high- resolution images to learn a mapping in a fully supervised manner. However, such image pairs are often not available in clinical practice. In this paper, we address super resolution problem in a real-world scenario using unpaired data and synthesize linearly \emph{eight times} higher resolved Micro-CT images of temporal bone structure embedded in the inner ear. We explore cycle-consistency generative adversarial networks for super-resolution and equip the model with Bayesian inference. We further introduce \emph{Hu Moments distance} as the evaluation metric to quantify the shape of the temporal bone. We evaluate our method on a public inner ear CT dataset and have seen both visual and quantitative improvement over state-of-the-art supervised deep-learning based methods. Further, we conduct a multi-rater visual evaluation experiment and find that three inner-ear researchers consistently rate our method highest quality scores among three methods. Furthermore, we are able to quantify uncertainty in the unpaired translation task and the uncertainty map can provide structural information of the temporal bone. 
%\vspace{-0.2cm}
\end{abstract}

%%%%%%%%%%%%%%%%%%%%%%%%%%%%%%%%%%%%%%%%%%%%%%%%%%%%%%%%%%%%%%%%%%%%%%%%%%%%%%%%
\vspace{-0.3cm}
\section{INTRODUCTION}
\vspace{-0.2cm}
High-quality image guided surgery has allowed clinicians to develop safe and less invasive surgical procedures despite the large inter-individual anatomical variability. Clinical computed tomography is the common imaging technique for this purpose. However, when delicate structures need to be resolved, for example the insertion of cochlear implant electrodes, a much higher resolution would be desirable. Micro computed tomography (micro-CT) can provide high resolution (HR) images that captures much more details of the human cochlear anatomy, however, it cannot be applied clinically due to the high amount of radiation dose required and due to the limited probe size which fits into the scanners. It is therefore highly clinically relevant to estimate detailed anatomical structures from low-resolution CT using super resolution (SR) techniques.
\\
\textbf{\emph{Related Work}}. Deep learning methods have been introduced to build a mapping function from LR patches to HR ones to solve various SR tasks \cite{oktay2016multi,park2018computed,tanno2017bayesian} and achieve superior results over traditional SR methods \cite{bahrami20177t,shi2013cardiac,alexander2017image}. 
Most of the deep learning-based methods rely on pairs of LR-HR images to train in a fully supervised manner. By downsampling the HR images to generate spatially-aligned LR ones, the paired images for training the deep networks can be easily obtained. 
However, this strategy fails to introduce real-world characteristics from the LR domain and often struggles to generalize when given real low-resolution images. Consequently, the models trained on simulated data become less effective if there exists a large domain shift between LR and HR domains. 
Especially in the SR task from clinical CT (LR) to Micro-CT (HR), noise, voxel size and sensor types result in large domain gap which make it extremely difficult to generate realistic paired LR-HR images. 
Recently, generative adversarial networks (GAN) and their extensions \cite{goodfellow2016nips,zhu2017unpaired,choi2018stargan} have been proposed to learn the data distribution from either paired or unpaired datasets. 
Although unpaired image-to-image translation using GAN was recently explored in medical imaging \cite{li2019diamondgan,yurt2019mustgan}, it cannot be directly applied to SR tasks because voxel sizes from the two domain are different. 
This work explores GAN for \emph{real-world} CT SR tasks and quantify model uncertainty in unpaired image translation. 
%\vspace{-0.4cm}
\\
\textbf{\emph{Contributions}}.
1) We propose a GAN-based approach to solve the unpaired super resolution problem in a clinical scenario and incorporate Bayesian inference to quantify model uncertainty to guarantee applicability in a translation to real world problems. 
2) We present qualitative and quantitative results on synthesis of Micro-CT, showing the superiority of our method over existing solutions. 
3) We report the results of multi-rater visual rating experiments, performed by three experienced inner-ear researchers who confirm the superior quality of generated high-resolution images while outperforming supervised deep learning approaches. 

%%%%%%%%%%%%%
\begin{figure}
	\begin{center}
	\vspace{0.2cm}
		\includegraphics[width=0.98\textwidth,height=0.61\textwidth]{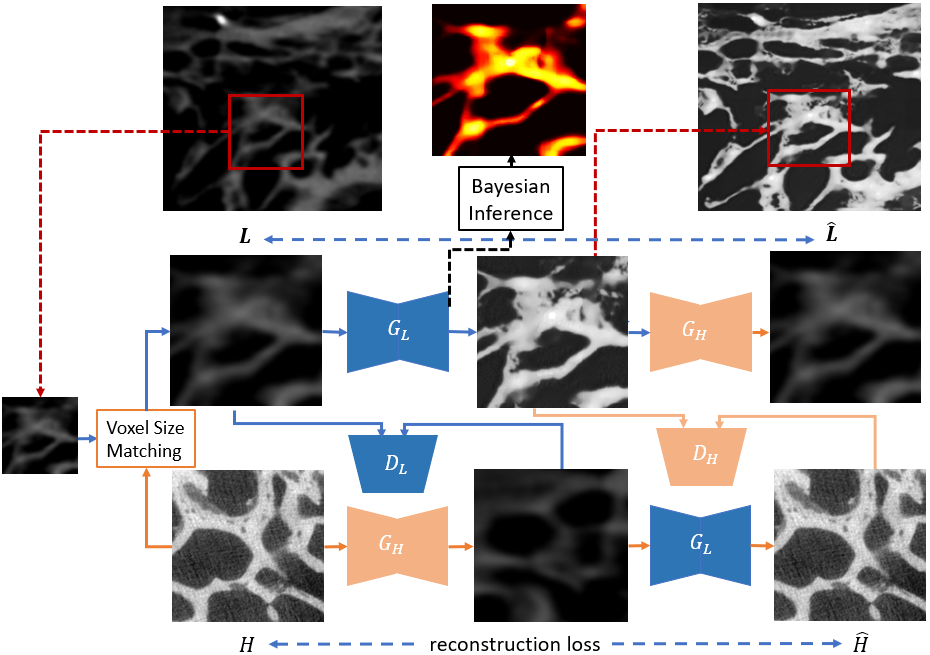}
	\end{center}
	\vspace{-0.3cm}
    	\caption{Method overview. We upsample the input LR images by bicubic interpolation to match the voxel size. Then we sample local patches from these upsampled slices of CT scans. Then the unpaired LR and HR image patches are fed into a cycle-consistency generative adversarial network to perform the super resolution task. Finally Bayesian inference is performed to estimate the model uncertainty given a patch.}
	\label{fig:overview} 
\end{figure}
%%%%%%%%%%%%%%%%%%%%%%%%%%%%%%%%%%%%%%%%%%%%%%%%%%%%%%%%%%%%%%%%%

\vspace{-0.3cm}
\section{Methods}
\vspace{-0.2cm}
Our super resolution approach includes two main components:  1) Unpaired image-to-image translation network and 2) Bayesian inference as shown in Figure  \ref{fig:overview}. 
~\\
~\\
 \textbf{\emph{Problem definition}}.  \label{problem_de}
The goal of the super resolution task is to map a low-resolution (LR) domain to a high-resolution (HR) domain.
Let $\mathcal{L}$ denote a LR image space and $\mathcal{H}$ a HR image space. 
We assume that $\mathcal{L}$ and $\mathcal{H}$ are not paired in a common clinical setting. 
Considering the complexity of imaging datasets, we claim that it is difficult to introduce LR domain characteristics in medical datasets by traditional techniques such as, interpolation and noise injection \cite{wang2019deep}, which restrict the applicability of SR techniques.   
Thus we aim to learn a direct mapping from LR domain to HR domain $\Phi $ : $\mathcal{L} \rightarrow \mathcal{H}$ by learning the individual underlying distributions. 
~\\
~\\
%\vspace{-0.2cm}
\textbf{\emph{Patch-Based Unpaired Image Translation}}.
The super-resolution task includes two aspects: 1) Structure enhancement and 2) Image style transfer. 
We deploy generative adversarial networks to learn the individual underlying distributions including generator networks and discriminator networks. 
Since it is an unpaired image-to-image translation task, we adapt a cycle-consistency GAN inspired by \cite{zhu2017unpaired} to handle the unpaired issue. This includes a pair of generators $\{G_{L}, G_{H}\}$ and a pair of domain discriminators $\{D_{L}, D_{H}\}$. As illustrated in Figure \ref{fig:overview}, generator $G_{L}$ aims to translate patches from LR to HR while generator $G_{H}$ aims to translate from HR to LR. Discriminators $\{D_{L}, D_{H}\}$ are trained to distinguish if the generated patches are real or fake in the two domains respectively. In a min-max game, the generators try to fool the discriminators by good image translation. 
The detailed training losses are as follows:
\\
\emph{Self-reconstruction loss:}
\vspace{-0.2cm}
\begin{equation}
\begin{aligned}
	\mathcal{L}_{rec} = \mathbb{E}_{x \sim \mathcal{L}}[||x-G_{H}(G_{L}(x))||_{1}] + \\ 
	\mathbb{E}_{y \sim \mathcal{H}}[||y-G_{L}(G_{H}(y))||_{1}]
D_{L}(G_{H}(y)))]\}
    \end{aligned}
\end{equation}
\\
\emph{Adversarial loss:}
\vspace{-0.2cm}
\begin{equation} \label{equation_1}
\begin{aligned}
\mathcal{L}_{adv} = {}&\mathbb{E}_{x \sim \mathcal{L}, y \sim \mathcal{H}}\{log~[D_{L}(x)\cdot D_{L}(y)]\} + \\
                        &\mathbb{E}_{x \sim \mathcal{L}, y \sim \mathcal{H}}\{log~[(1 - D_{H}(G_{L}(x)))]+ \\
                        &\mathbb{E}_{x \sim \mathcal{L}, y \sim \mathcal{H}}\{log~[(1 - D_{L}(G_{H}(y)))]\}
\end{aligned}
%\vspace{-0.2cm}
\end{equation}

The full objective function of our framework is:
\begin{equation}
	\mathcal{L}_{total} =  \lambda_{rec}\mathcal{L}_{rec} + \lambda_{adv}\mathcal{L}_{adv}
\end{equation}

where $\lambda_{rec}$ and $\lambda_{adv}$ control the importance of self-reconstruction and adversarial training.  
\\
~\\
\emph{Whole-slice reconstruction}: Since the method is patch-based considering the computation complexity, we further reconstruct the whole slice from individual generated image patches. As stitching patches leads to a grid effect, we eliminate this effect by first performing histogram matching for each generated patch using the corresponding LR patches and then do median filtering as a post-processing step to remove the grid effect while preserving the structural information. 
~\\
~\\
\textbf{\emph{Bayesian Inference}}.
To estimate the model uncertainty, we employ Monte Carlo dropout \cite{gal2015dropout} to run multiple forward passes through the model using a dropout layer. To derive the uncertainty for one patch $x\in \mathcal{L}$, we collect the predictions of $T$ inferences with different dropout masks. Let $G_{L}^{d_{i}}$ denote the generator $G_L$ with dropout mask $d_{i}$. We obtain a sample of the possible model outputs distribution for a patch $x$ as \{$G_{L}^{d_{0}}(x), ..., G_{L}^{d_{T}(x)}$\}. We compute the mean and variance of $x$, corresponding to the mean of model posterior distribution and the estimation of model uncertainty. The predictive posterior mean $p$ and uncertainty $c$ are formulated as:
\begin{equation}
	p =  \frac{1}{T} \sum_{i=0}^{T}G_{L}^{d_{i}} ;~~c=  \frac{1}{T} \sum_{i=0}^{T}[G_{L}^{d_{i}}-p]^2 
\end{equation}
\textbf{\emph{Implementation}}.
The voxel size matching between two scanners is performed by bicubic interpolation. 
Each generator network contains two convolutions with a stride of two, nine residual blocks \cite{he2016deep} and two convolutions with a stride of 1/2. Six residual blocks were used for the input with a size of $h \times w \times 1$, where $h$ and $w$ are the height and width of the input images respectively. Discriminators are convolutional neural networks for classification of real and generated images (i.e. 'real' or 'fake'). We leverage PatchGANs \cite{johnson2016perceptual} to enforce the discriminator network to classify the local patch features to real or fake, instead of the global image.
For training, we use the Adam optimizer with a batch size of 5, a learning rate of 0.0001 and a number of 50 epochs. 
In all experiments, we set the hyper-parameters $\lambda_{adv}$ and $\lambda_{rec}$ to 1 and 10 respectively. 
 Experiments are run on one Nvidia Titan Xp GPU. The training time is around eight hours. The inference stage takes 2 mins for 200 patches in one slice. 
\\
~\\
%\vspace{-0.2cm}
\textbf{\emph{Visual Evaluation Protocol and Evaluation Metrics}}. \label{protocol}
Due to lack of paired images, quantitative analysis of images cannot be directly performed using traditional metrics such as PSNR and SSIM \cite{welander2018generative}. Additionally these metrics are not able to capture and evaluate useful information such as preserving the clinically relevant structures. In order to get a qualitative assessment of the images generated by our approach, we design a visual rating protocol to compare the images generated by our approach with those generated by two state-of-the-art methods (i.e. U-Net \cite{ronneberger2015u} and SR-GAN \cite{ledig2017photo}). In each trial, the otolaryngology researchers were presented with three sets of images, with each set consisting of an original LR CT image on the left and the synthetic image generated by U-Net, SR-GAN or our approach. The rater does not know which method is used to prevent any bias. The rater is asked to rate on the following criterion: a) good resolution and low noise level and b) good shape and structural information. The otolaryngology researchers were asked to rate on a six star scale where `one` represents `poor` and `six` represents `excellent`. 

To quantitatively compare our method with others in the unpaired setting, we introduce \emph{Hu Moments distance} \cite{vzunic2010hu} as an evaluation metric on shape similarity due to its rotation- and scale- invariance in shape encoding. 
However, since the generated images and HR image are not aligned, it is not possible to directly measure the shape similarity. Intuitively, the shape of the structures from real and fake images are expected to be similar. We propose \emph{minimum Hu Moments distance} (m-HuM) to handle the similarity evaluation for unpaired slices. Given two sets of unpaired data $\mathcal{X}$ and $\mathcal{Y}$, we take the minimum HuM distance for all possible combinations of the elements between the two sets:
  \vspace{-0.2cm}
\begin{equation} \label{equation_6}
\emph{m-HuM} = \min\{HuM(x, y)| x\in \mathcal{X}, y\in \mathcal{Y}\}
\end{equation}

\section{Experiments}
 \vspace{-0.2cm}
\textbf{\emph{Dataset and Evaluation Metric}}.
%%%%%%%%%%%%%%%%%%
 We evaluate the proposed method on a public human inner ear dataset \cite{gerber2017multiscale} including two diverse CT scanners as shown in Table \ref{tab:datasets}. Specifically we perform linearly \emph{eight times} super resolution from clinical cone beam CT to the state-of-the-art Micro CT. 
For pre-processing, all the low-resolution scans are matched in voxel size using bicubic interpolation and for each slice, we crop the background to obtain a region of interest by thresholding and using a bounding box. Then they are normalized using $z-score$ normalization and the intensity is then normalized to the range of [0, 1] to meet the requirement of the network output. We then extract patches with a size of $256 \times 256$ from whole slices with a stride of 128 pixels resulting in a training set of ~5K patches. We use six LR scans from subset `AL\_1` and six HR ones from subset `BH\_1` as a training set. Also, we use one paired scan (with same patient ID) from `BL\_1` and `AH\_1` as a test set. Note that even though the two scans are from the same patient, it is extremely difficult to align them spatially.

\begin{table*}[htpb]

  \caption[table: Data Properties]{Data characteristics of a subset of the public inner ear dataset including one Micro-CT and one clinical CT. We used five cone beam CT scans and five Micro-CT scans for the unpaired training and the rest for testing.}\label{tab:datasets}
  \centering
  \vspace{-0.4cm}
   \begin{tabular}{l l l c c c }
    \hline
      ~~Scanner~~&~~~Voxel Size (mm$^3$)~~~&~~Volume Size~~& Subjects\\
    \hline
      Micro-CT& 0.018 $\times$ 0.018 $\times$ 0.018 &~$1828\times1828\times1828$ &7 \\
      cone beam CT &$~0.15\times0.15\times0.15$ &~~$668\times668\times668~~$ &7\\
     \hline
  \end{tabular}
  \vspace{-0.2cm}
\end{table*}
%%%%%%%%%%%%%%%%%

%%
%%%%%%%%%%%%%%%%%%%%%%%%%%%%%%%%%%%%%%%%%%%%%%%%%%%%%%%%%%%%%%%%%%%%%%%%%%%%%%
%\vspace{-0.1cm}
\begin{figure*}[t]
	\begin{center}
		\includegraphics[width=0.88\textwidth,height=0.44\textwidth]{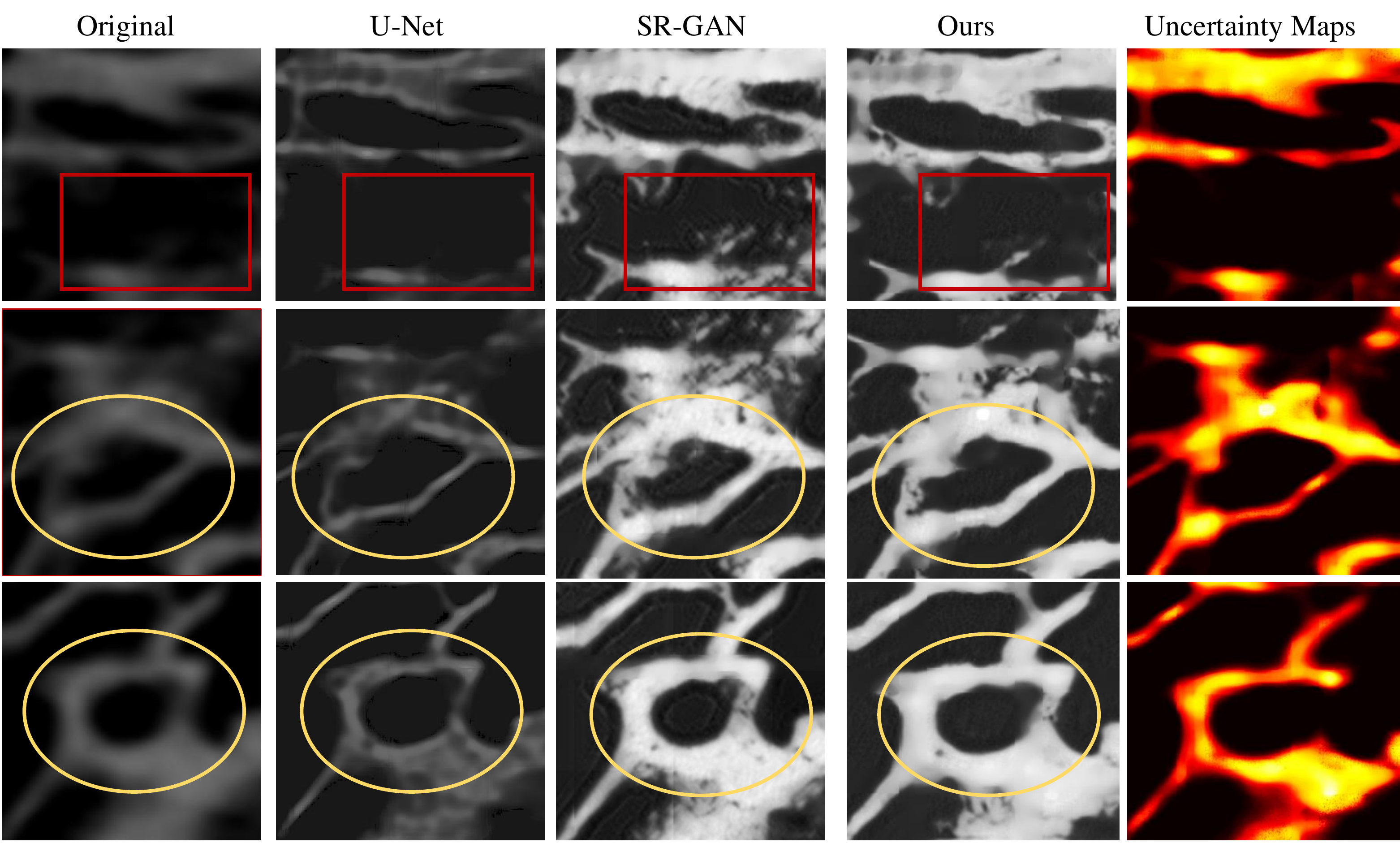}
	\end{center}
	\vspace{-0.6cm}
    	\caption{Qualitative results of three approaches. The region marked with red rectangle in the first row shows the artefacts generated by SR-GAN. The region marked with yellow circle shows that our approach generates good shape and local structure with less noise. The uncertainty maps highlight the structure of the temporal bone and can serve as a rough segmentation map.}
	\label{fig:results_SOTA} 
\end{figure*}
%%%%%%%%%%%%%%%%%%%%%%%%%%%%%%%%%%%%%%%%%%%%%%%%%%%%%%%%%%%%%%%%%%%%%%%%%%%%%%
~\\
~\\
\textbf{\emph{Qualitative and Quantitative Results}}.
We firstly evaluate our method on the test set with state-of-the-art methods U-Net \cite{ronneberger2015u} and SR-GAN \cite{liu2017unsupervised}. Note that both of them are supervised-learning methods and require paired data for training. 
For these two methods, we synthesize LR images by bicubic interpolation considering the voxel size difference and inject Gaussian noise as used in previous SR tasks \cite{wang2019deep}. 
As shown in Fig. \ref{fig:results_SOTA}, visually our method enhances the temporal bone structure and preserves good-quality shape information compared to other supervised approaches. We also observe that SR-GAN generates some artefacts in both background and structure boundary by hallucinating noisy patches which are dominated by continuous partial volume effects that are of no relevance for the bony structures, while U-NET preserves the shape structure but struggles to enhance the local structure. Although SR-GAN uses a GAN component to enhance the image quality, it does not guarantee the model can be generalized to real data.
We further perform $T$ forward passes through the model using a dropout layer to collect the predictions. In practice, we set $N = 100$ considering the computation complexity.
The heatmaps in Fig. \ref{fig:results_SOTA} depict the uncertainty obtained for the input patches. 
We observe that uncertainty maps match well with the temporal bone structure and thus they can serve as segmentation maps to interpret the structure in LR images. 

Table. \ref{tab:results_SOTA} shows the quantitative comparison of three methods using \emph{m-HuM}. For each whole slice, we calculate the \emph{m-HuM} by comparing it to the whole test set. The results of all testing slices are averaged. It demonstrates that our method outperforms U-Net and SR-GAN in terms of preserving the shape of temporal bone structure. 

\begin{table}[htpb]
  \caption[table: Data Properties]{Comparison with state-of-the-art supervised methods using \emph{minimum Hu Moments distance}. $\downarrow$ indicates that a lower value represents a higher shape similarity.}
  \label{tab:results_SOTA}
  \centering
   \begin{tabular}{c c c c c }
    \hline 
     Methods&~~U-Net \cite{ronneberger2015u}~~& SR-GAN \cite{ledig2017photo} &~Ours~\\
    \hline
      \emph{m-HuM} $\downarrow$  & 3.12 & 1.29  & \textbf{0.28}\\
     \hline
  \end{tabular}
  \vspace{-0.2cm}
\end{table}

%%%%%%%%%%%%%%%%%%%%%%%%%%%%%%%%%%%%%%%%%%%%%%%%%%%%%%%%%%%%%%%%%%%%%%%%%%%%%%

\makeatletter
\def\thickhline{%
  \noalign{\ifnum0=`}\fi\hrule \@height \thickarrayrulewidth \futurelet
   \reserved@a\@xthickhline}
\def\@xthickhline{\ifx\reserved@a\thickhline
               \vskip\doublerulesep
               \vskip-\thickarrayrulewidth
             \fi
      \ifnum0=`{\fi}}
\makeatother

\newlength{\thickarrayrulewidth}
\setlength{\thickarrayrulewidth}{2\arrayrulewidth}

\begin{figure}[t]
	\begin{center}
		\includegraphics[width=1\textwidth,height=0.485\textwidth]{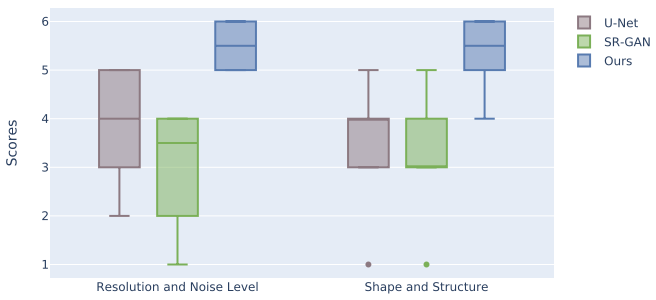}
	\end{center}
%	\vspace{-0.2cm}
    	\caption{Rating results by three inner-ear researchers. Our method generates good-quality images and significantly outperforms other supervised methods, benefiting from learning distributions on real data using GAN. }
	\label{fig:rating_results} 
\end{figure}

%%%%%%%%%%%%%%%%%%%%%%%%%%%%%%%%%%%%%%%%%%%%%%%%%%%%%%%%%%%%%%%%%%%%%%%%%%%%%%
%\vspace{-0.3cm}
~\\
\textbf{\emph{Visual Evaluations}}.
In order to assess how our approach generates images with clinical relevance, three raters with a mean of 7+ years of experience rated the image quality based on the criterion mentioned in Section \ref{protocol}. 
Results on the test set for the different metrics (i.e. resolution, noise level and preservation of shape and structure) are presented in the boxplots in Fig. \ref{fig:rating_results}.
It shows that our method outperforms U-Net and SR-GAN in both metrics. We observe that U-Net produces higher quality images than SR-GAN in terms of resolution and noise level. 
For each sample, we conduct Wilcoxon rank-sum test on the paired rating scores on two testing scans from three raters to compare the methods. Results show that the pair of metric ratings on our approach with other two methods are significantly different (with p-values $<$ 0.0001). This demonstrates that our method significantly outperforms U-Net and SR-GAN in image quality. As claimed in Section \ref{problem_de}, our method benefits from learning the real distribution of LR domain instead of introducing artificial factors. This strategy strengthens its applicability to real data.  

%\vspace{-0.2cm}
%%%%%%%%%%%%%
\vspace{-0.2cm}
%%%%%%%%%%%%%%%%%%%%%%%%%%%%%%%%%%%%%%%%%%%%%%%%%%%%%%%%%%%%%%%%%

%%%%%%%%%%%%%%%%%%%%%%%%%%%%%%%%%%%%%%%%%%%%%%%%%%%%%%%%%%%%%%%%%%%%%%%%%%%%%%
\vspace{-0.2cm}
\begin{figure*}[t]
	\begin{center}
		\includegraphics[width=0.88\textwidth,height=0.465\textwidth]{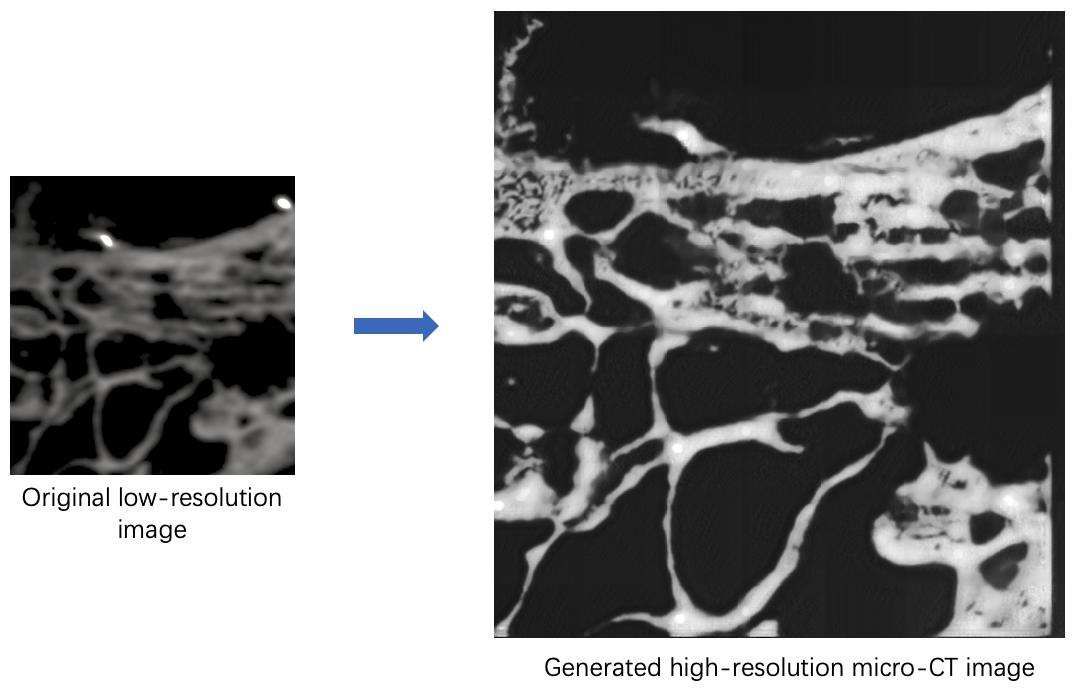}
	\end{center}
	\vspace{-0.5cm}
    	\caption{One synthetic high-resolution micro-CT slice (right) generated from a clinical CT image (left) of one testing subject. Notably it is very challenging to match the corresponding ground truth high-resolution image due to the random 3D orientation.}
	\label{fig:results_samples} 
\end{figure*}
%%%%%%%%%%%%%%%%%%%%%%%%%%%%%%%%%%%%%%%%%%%%%%%%%%%%%%%%%%%%%%%%%%%%%%%%%%%%%%

%\vspace{-0.1cm}
\section{Summary and Conclusion}
\vspace{-0.1cm}
This work introduces a super resolution approach integrating generative adversarial network and Bayesian inference. For the first time, we address Micro-CT image synthesis in a real-world scenario using only unpaired data. This approach is potentially helpful for image-guided surgery of cochlear interventions. 
As a promising approach, we will investigate its usage for safer and less invasive surgical procedures. %In addition, the uncertainty maps of our approach highlight the temporal bone structure and can serve as a basis for segmentation. 
\vspace{-0.3cm}
\section{Compliance with Ethical Standards}
\vspace{-0.2cm}
The study was carried out in accordance with \emph{the Declaration of Helsinki}.
\section{Acknowledgement}
\vspace{-0.3cm}
The project was funded in part by the German Research Foundation (DFG HE6713/2-1).
\\
~\\
{\footnotesize
\bibliographystyle{ieee}
\bibliography{egbib}
}

\end{document}